\documentclass{article}
\usepackage{kmn,latexsym,subfigure,epsfig,graphics}
\begin{document}

\title[Thermal Desorption of Water-Ice in the ISM]{Thermal Desorption of Water-Ice in the Interstellar Medium}
\author[Fraser et al.]{Helen J. Fraser$^{1}$\thanks{The Raymond \& Beverly Sackler Laboratory for Astrophysics, Sterrewacht Leiden, Leiden University, Postbus 9513, 2300 RA Leiden, The Netherlands.}\thanks{ Corresponding author: The Raymond \& Beverly Sackler Laboratory for Astrophysics, Sterrewacht Leiden, Leiden University, Postbus 9513, 2300 RA Leiden, The Netherlands {fraser@strw.leidenuniv.nl}}, Mark P. Collings$^{1}$, Martin R.S. McCoustra$^{1}$ and David A. Williams$^{2}$ \\
$^{1}$School of Chemistry, University of Nottingham, University Park, Nottingham, NG7 2RD, U.K.\\
$^{2}$Department of Physics and Astronomy, University College London, Gower
Street, London, WC1E 6BT, U.K.\\
}
\maketitle

\begin{abstract}
Water (H$_{{\rm 2}}$O) ice is an important solid constituent of many
astrophysical environments. To comprehend the role of such ices in the
chemistry and evolution of dense molecular clouds and comets, it is
necessary to understand the freeze-out, potential surface reactivity, and
desorption mechanisms of such molecular systems. Consequently, there is a
real need from within the astronomical modelling community for accurate
empirical molecular data pertaining to these processes. Here we give the
first results of a laboratory programme to provide such data. Measurements
of the thermal desorption of H$_{{\rm 2}}$O ice, under interstellar
conditions, are presented. For ice deposited under conditions that
realistically mimic those in a dense molecular cloud, the thermal desorption
of thin films ($\approx$50 molecular layers) is found to occur with zero order
kinetics characterised by a surface binding energy, $E_{des}$, of 5773 $\pm
$60 K, and a pre-exponential factor, $A$, of 10$^{{\rm 3}{\rm 0}{\rm} {\rm \pm
}{\rm 2}}$ molecules cm$^{{\rm -} {\rm 2}}$ s$^{{\rm -} {\rm 1}}$. These
results imply that, in the dense interstellar medium, thermal desorption of
H$_{{\rm 2}}$O ice will occur at significantly higher temperatures than has
previously been assumed.

\end{abstract}


\section{Introduction}

It has become clear over the last decade that purely gas phase schemes
cannot account for the variety and richness of chemistry occurring in the
interstellar medium (ISM), particularly in denser regions of molecular
clouds where star formation is occurring. In such environments gas-grain
interactions\textbf{} must also play a key role (Williams 1998). Within the
lifetime of a dense molecular cloud (on the order of 10$^{{\rm 6}}$ yr), the
grains accrete icy mantles as atomic and molecular species freeze-out onto
the grains. The ices are processed further by cosmic ray impacts, UV
photolysis, and shocks. Consequently, the chemical composition of these icy
mantles is significantly different from that of the local gas phase
environment (Rawlings 2000). During cloud collapse and stellar formation the
grains are reheated, and molecules are recycled from the solid state into
the gas phase environment. In regions where the H$_{{\rm 2}}$O
concentrations are depleted it is assumed that the species is still locked
in the icy mantles on the grains, and consequently, that the temperature is
below the sublimation temperature of the tracer gas (van Dishoeck \& van der
Tak 2000). When evaluating these environments it is clearly necessary to
understand the sublimation behaviour of the ice, and its thermal stability.
The surface binding energy, \textit{${\rm E}$}$_{des}$, and sublimation rates of molecular
ices may only be obtained empirically. When combined, these data can be used
to determine molecular residence times on the ice surface, in terms of a
population half-life as a function of temperature. These data are also
applicable in other astronomical environments, such as cometary nuclei and
comae, planetary surfaces and satellites.

\bigskip

H$_{{\rm 2}}$O is the most abundant molecular ice in dense ISM regions
(Ehrenfreund \& Schutte 2000). The H$_{{\rm 2}}$O ice band at 3.07 $\mu $m
is detected on lines of sight towards reddened stars that have a visual
extinction above a critical value (Whittet 1993) and is typically one of the
strongest bands in interstellar IR spectra. The abundance and morphology of
H$_{{\rm 2}}$O ice in the ISM is generally inferred from the intensity and
line profile of the 3.07 $\mu $m band. At low pressures and temperatures,
such as those in the ISM, the H$_{{\rm 2}}$O ice can exist in both amorphous
and crystalline forms. To further complicate the picture, the amorphous ice
can exist in both high-density (1.1 g cm$^{{\rm -} {\rm 3}}$) and
low-density (0.94 g cm$^{{\rm -} {\rm 3}}$) forms, reflecting differences in
the porosity of the material, with a phase change from former to latter
occurring irreversibly between 38 -- 80 K (Jenniskens et al$.$, 1995). It is
generally assumed that the dominant morphology of icy mantles on
interstellar grains resembles that of high density amorphous ice, with other
`impurity' molecules trapped in the H$_{{\rm 2}}$O ice matrix (Jenniskens et
al$.$ 1995). This is evident in comparisons of ISO SWS and ground based
observations with laboratory spectra of mixed ice analogues (Ehrenfreund \&
Schutte 2000 and references therein). However, laboratory spectra of pure
H$_{{\rm 2}}$O ices show clear differences in the line profile of the 3.07
$\mu $m feature associated with different ice morphologies and temperatures
(Jenniskens et al$.$ 1997). It is therefore likely that some H$_{{\rm 2}}$O
ices have undergone `mild processing' and re-cooling, (such as gentle
heating below 80 -- 100 K, or rapid localised heating due to shocks), and
will therefore also exist in low-density amorphous and crystalline forms
(Jenniskens \& Blake, 1994, 1996). The physical properties of the H$_{{\rm
2}}$O ice, such as density, conductivity, vapour-pressure, and sublimation
rate, are dictated by its structure. Significant differences are expected
between the surface chemistry and bulk behaviour of the ice phases. In
addition, the physical and chemical properties of the ice may also be
affected by way the ice film is deposited, its lifetime, and processing
prior to thermal desorption (Sack \& Baragiola 1993).

\bigskip

This paper is the first in a series of papers reporting new experimental
results on gas-surface interactions occurring under conditions resembling
those in the ISM and star-forming regions. With the aid of a purpose-built
instrument, we are now able, for the first time, to provide the
astrochemical community with experimental results addressing not only
desorption energies, but also sticking probabilities and desorption
mechanisms. Here we present the results of laboratory studies to determine
the surface binding energies on, and sublimation rates of H$_{{\rm 2}}$O ice
under interstellar conditions. Previous studies, where the surface binding
energies were determined from spectroscopic data, suggested that the surface
binding energy of H$_{{\rm 2}}$O on H$_{{\rm 2}}$O ice was 4815 $\pm $15 K
or 5070 $\pm $50 K, for unannealed and annealed ice samples respectively
(Sandford \& Allamandola 1988). It was also presumed that the desorption
kinetics of the system were first order. These data have been used
extensively in the astronomy literature, where it is now widely reported
that under interstellar conditions H$_{{\rm 2}}$O can only exist in the
solid phase below about 100 K (Hasegawa \& Herbst 1993, Sandford \&
Allamandola 1993, Ehrenfreund \& Schutte 2000). The direct techniques used
to determine the surface binding energy in this work are described in \S 2.
The results are detailed in \S 3. In these experiments the thermal
desorption of thin H$_{{\rm 2}}$O ice films is found to occur with zero
order kinetics characterised by a surface binding energy, $E_{des}$, of 5773
$\pm $60 K, and a pre-exponential factor, A, of 10$^{{\rm 3}{\rm 0}{\rm
}{\rm \pm} {\rm 2}}$ molecules cm$^{{\rm -} {\rm 2}}$ s$^{{\rm -} {\rm 1}}$.
In \S 4, a comparison is made with previous results, from both thermodynamic
and spectroscopic techniques and the astrophysical implications of this work
are discussed, focusing on applications associated with dense molecular
clouds. The conclusions are summarised in \S 5.

\section{EXPERIMENTAL METHOD}

The Nottingham Surface Astrophysics Experiment (NoSAE) was designed to
measure surface binding energies and sticking probabilities empirically and
accurately, recreating the harsh interstellar environment under controlled
laboratory conditions. A much fuller description than can be given here is
to be found elsewhere (Fraser, Collings \& McCoustra 2001). Basically, the
experiment consists of a stainless steel ultrahigh vacuum (UHV) chamber with
an operating base pressure of 6$\times $10$^{{\rm -} {\rm 1}{\rm 1}}$ Torr.
The primary constituent (>90\% ) of this vacuum is H$_{{\rm 2}}$, providing
very similar conditions with the experiment chamber as those found in the
ISM. The chamber is equipped with an effusive gas deposition system,
quadrupole mass spectrometer (QMS), FTIR spectrometer and a Quartz Crystal
Microbalance (QCM). The QMS is used for temperature programmed desorption
(TPD) experiments (see \S 2.2), the FTIR for reflection-absorption infrared
spectroscopy (RAIRS), and the QCM for thin film mass determination. The
grain mimic is an uncharacterised gold film surface, suspended at the end of
a cryogenic cold finger, capable of reaching 7 K. The sample is radiatively
heated from the reverse side by a halogen bulb. The sample temperature is
measured using two E-type (Ni-Cr alloy vs. Cu-Ni alloy) and one KP-type (Au
including 0.7\% Fe vs. Chromel) thermocouples, positioned on or near to the
sample. The temperature can be controlled to better than 0.5 K, and measured
accurately within 0.25 K. The upper temperature range of the system is 350
K. The whole system is linked via DAQ, GPIB, and RS232 communications to a
pair of PC's, which are used to co-ordinate and control data acquisition
from various instruments on the experiment.

\subsection{Deposition}

H$_{{\rm 2}}$O was obtained \textit{in situ} from a liquid H$_{{\rm 2}}$O sample that had
been deionized, filtered for organic and inorganic solutes, and subjected to
three freeze-thaw cycles under a vacuum of better than 1$\times $10$^{{\rm -
}{\rm 7}}$ Torr. H$_{{\rm 2}}$O ice films were deposited at a rate of \textit{ca.}
10$^{{\rm 1}{\rm 0}}$ molecules s$^{{\rm -} {\rm 1}}$ on the gold film
substrate held at a temperature of 10 K, from a quasi-effusive molecular
beam of H$_{{\rm 2}}$O, directed at 5$^{{\rm o}}$ to the surface normal. At
such low temperatures and slow deposition rates ballistic deposition results
in the formation of high-density amorphous ice. Exposure of the substrate to
gaseous H$_{{\rm 2}}$O was varied between 1 and 100 Langmuir (L), (1
Langmuir is equivalent to 1$\times $10$^{{\rm -} {\rm 6}}$ Torr s), and film
mass was estimated using the QCM. An exposure of 20 L corresponded to a film
mass of \textit{ca.} 3.8$\times $10$^{{\rm -} {\rm 7}}$ g, which equates to a surface
number density of H$_{{\rm 2}}$O of 2.44$\times $10$^{{\rm 1}{\rm 6}}$
molecules cm$^{{\rm -} {\rm 2}}$, or a film thickness of \textit{ca.} 0.03 $\mu $m, when
proper account of the QCM sensor area is made.

\subsection{Temperature programmed desorption}

The sublimation characteristics of the H$_{{\rm 2}}$O films were measured by
temperature programmed desorption (TPD). In this method a linear temperature
ramp is applied to the sample and the rate of desorption is measured by
monitoring the amount of adsorbate that desorbs into the gas phase as a
function of temperature. Desorption is an activated process which takes
place at a rate given by

\begin{equation}
\label{eq1}
R = - {\frac{{dN_{S}} }{{dt}}} = k_{d} N_{S}^{m}
\end{equation}

\noindent
where $N_{S} $is the number of adsorbed molecules on the surface in molecules
cm$^{{\rm -} {\rm 2}}$, $m$ is the order of the reaction and $k_{d}$ is the rate
constant, given by

\begin{equation}
\label{eq2}
k_{d} = A\exp \left( { - {\frac{{E_{des}} }{{k_{B} T}}}} \right)
\end{equation}

\noindent
where $A$ is the pre-exponential factor, $E_{des}$ is the binding energy of a
molecule on the ice surface in J, $k_{B}$ is the Boltzmann constant and $T$ is
the ice temperature in K. For convenience, in this paper the surface binding
energies are expressed as $E_{des}/k_{B}$, i.e. in K, and the units of $A$ are
determined from $k_{d}$, i.e. for $m$ = 1, $A$ is expressed in s$^{{\rm -} {\rm
1}}$, and where $m$ = 0, $A$ is expressed in molecules cm$^{{\rm -} {\rm 2}}$
s$^{{\rm -} {\rm 1}}$. Equation (\ref{eq1}) can be rewritten, to reflect the TPD
signal that is actually measured during the experiment, i.e.

\begin{equation}
\label{eq3}
 - {\frac{{dN_{S}} }{{dT}}} = k_{d} {\frac{{N^{m}}}{{\beta} }}
\end{equation}

\noindent
where $\beta $ is the heating rate, \textit{dT/dt}. The recorded TPD signal peaks at some
temperature maximum, $T_{d}$, which corresponds to the point at which the
desorption rate from the surface is at a maximum, i.e.
($d^{2}N_{S}$\textit{/dt}$^{2}$ = 0). The TPD spectra are collected at increasing
initial surface coverages, using the same linear heating ramp in each case.
To a first approximation, the surface binding energy, $E_{des}$, can then be
calculated directly by substituting for $k_{d{\rm} }$in equation (\ref{eq3}) (using
equation (\ref{eq2})), differentiating and equating to zero

\begin{equation}
\label{eq4}
{\frac{{E_{des}} }{{k_{B} T_{d}^{2}} }} = {\frac{{A}}{{\beta} }}mN_{S}^{m -
1} \exp \left( { - {\frac{{E_{des}} }{{k_{B} T_{d}} }}} \right)
\end{equation}

\noindent
provided the pre-exponential factor, $A$, is known. In a first order desorption
process, where $m$ = 1, $A$ is assumed to be around 10$^{{\rm 1}{\rm 2}}$ to
10$^{{\rm 1}{\rm 3}}$ s$^{{\rm -} {\rm 1}}$, approximately the vibrational
frequency of the weak bond between the adsorbate molecule and the surface.
It is also possible to evaluate the order of the reaction from the TPD peak
shape and peak maximum provided that the activation energy for desorption
and pre-exponential factor remain constant as a function of surface
coverage. In practice however, desorption may not occur in a single step,
the surface binding energy can vary across binding sites on the surface or
with surface coverage, and the pre-exponential factor, $A$, may differ by
several orders of magnitude. A more rigorous result can be obtained by
modelling the desorption system and using numerical fitting methods with a
number of TPD spectra to evaluate the values of $A$ and $E_{des}$ (see Woodruff
\& Delchar 1986, Attard \& Barnes 1998 for a more complete discussion).

In these experiments the ice layers were warmed at a rate of 0.02 K s$^{{\rm
-} {\rm 1}}$ and the QMS, tuned to mass 18 (H$_{{\rm 2}}$O$^{{\rm +} }$),
was used to monitor the gas phase composition during desorption from the
substrate. The QMS was operated in a line-of-sight configuration so that
only molecules originating from the sample surface produced a signal at the
detector (Fisher \& Jones 1999). Sample coverages of 1, 2, 5, 10, 20, and 50
L (sub-monolayer to multilayer) were used. In all cases, no desorption was
observed before 120 K and desorption was complete by 170 K.

\section{RESULTS}

\begin{figure}[!ht]
{\hbox to 0.45\textwidth{\epsfxsize=0.45\textwidth 
\epsfbox{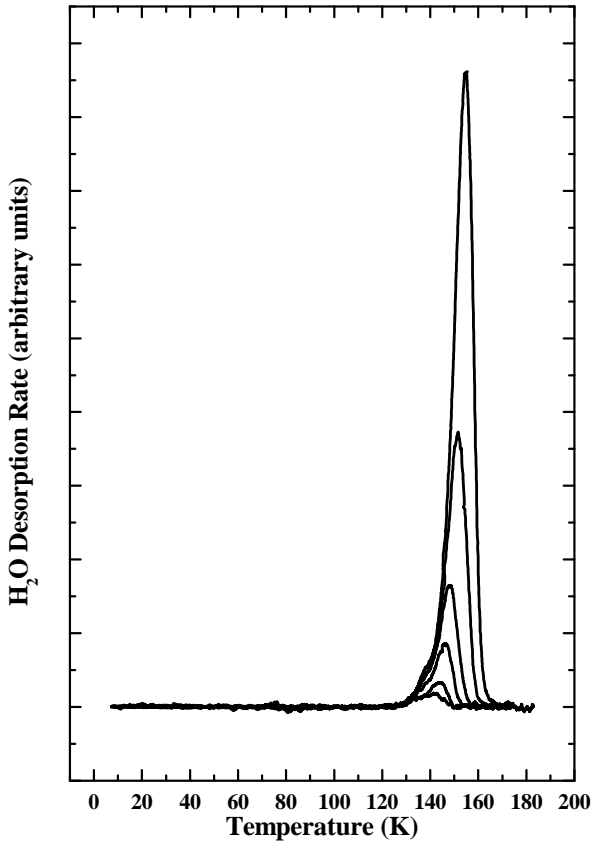}}}
{\caption{TPD spectra for H$_{{\rm 2}}$O ice films deposited at 10 K on a Au
substrate. Spectra are shown for surface exposures of 1, 2, 5, 10, 20, and
50 L. See text for details.}}
\end{figure}

Fig. 1 shows a sequence of TPD spectra for H$_{{\rm 2}}$O films deposited at
10 K, with increasing surface coverage. The spectra show the change in
intensity of the mass 18 (H$_{{\rm 2}}$O$^{{\rm +} }$) signal at the QMS as
the sample is slowly heated. An intense, asymmetric peak is observed at
\textit{ca.} 140 K, associated with the desorption of the bulk ice film from the Au
substrate. There is little or no change in the shape of the leading edge
peak with increasing exposure. Furthermore, the peak temperature,
$T_{d},$ increases monotonically with exposure. These observations are typical
of TPD spectra in which the desorption process follows zero order kinetics,
as illustrated by Fig. 2. The figure shows results from a series of (a)
zeroth and (b) first order TPD simulations (based on equation 3), at
coverages of 2, 5, 10, 20, 50 and 100 monolayers of H$_{{\rm 2}}$O, and
under conditions similar to those in this experiment. In the zeroth order
case (Fig. 2 (a)), the TPD curves are asymmetric, with coincident rising
edges, and the peak temperature, $T_{d}$, increases with exposure. This is
typical of multilayer desorption, where the intensity of the TPD spectrum
increases as more and more material is condensed onto the surface. The shift
in $T_{d}$\textit{} (the peak position) occurs because the desorption rate increases
exponentially with temperature, so the rate can increase indefinitely until
all the layers have been stripped away, and the TPD signal falls rapidly to
zero. The bonding between the layers resembles that in a condensed solid of
the adsorbate, so the desorption energy, $E_{des}$, and pre-exponential
factor, $A$, are the same for every layer. In contrast, the first order
desorption curves are symmetric about a single $T_{d{\rm} }$value,
independent of coverage (see Fig. 2 (b)). This closely resembles monolayer
desorption kinetics, when the nature of the bonding between the adsorbate
and the underlying substrate is clearly important. In a multilayer system
(e.g. icy mantles on dust grains, or the ice system studied here) the
monolayer is generally more tightly bound to the underlying substrate than
the subsequent adsorbate layers. Consequently, the monolayer desorption peak
appears at higher temperatures in the TPD spectrum than the multilayer
desorption, reflecting a different $E_{des}$ value. Its line shape profile
will also differ, reflecting both a different $A $value, and the different
reaction kinetics. A more complete discussion of this complex relationship
between TPD line shape profiles, binding energies, and desorption kinetics
can be found in Woodruff \& Delchar (1986) or Attard \& Barnes (1998).
However, from a comparison between the experimental TPD spectra presented in
Fig. 1 and the simulations in Fig. 2, it is reasonable to conclude that the
results of this experiment are indicative of zeroth order kinetics.
Furthermore, only a single peak is observed in the TPD spectra: even at low
coverage it is not possible to deconvolve a monolayer desorption peak from
the multilayer peak. Therefore no evidence exists to suggest that the
desorption process is coverage dependant, nor that it is possible to
distinguish between the multilayer and monolayer desorption processes. These
observations are entirely consistent with previous reports on the TPD of ice
from other hydrophobic, metal substrates (Kay et al. 1989; Dohn\'{a}lek et
al$.$1999, 2000).

\begin{figure}[!ht]
{\hbox to 0.45\textwidth{\epsfxsize=0.45\textwidth
\epsfbox{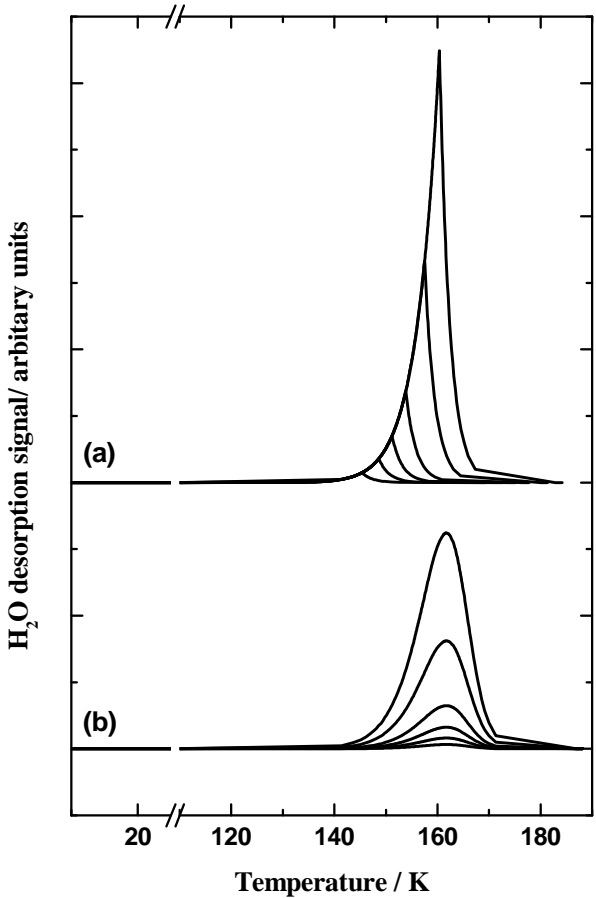}}}
{\caption{TPD spectra from a series of (a) zeroth ($A$ = 10$^{{\rm 3}{\rm 0}}$
molecules cm$^{{\rm -} {\rm 2}}$ s$^{{\rm -} {\rm 1}}$) and (b) first order
($A $= 10$^{{\rm 1}{\rm 3}}$ s$^{{\rm -} {\rm 1}}$) simulations (based on
equation 3), at coverages of 2, 5, 10, 20, 50 and 100 monolayers of H$_{{\rm
2}}$O. In each case, the desorption energy, $E_{des}$, was fixed at 5773 K,
and the heating ramp, \textit{$\beta $}, was 0.01 Ks$^{{\rm -} {\rm 1}}$. }}
\end{figure}

\begin{figure}[!hb]
{\hbox to 0.45\textwidth{\epsfxsize=0.45\textwidth
\epsfbox{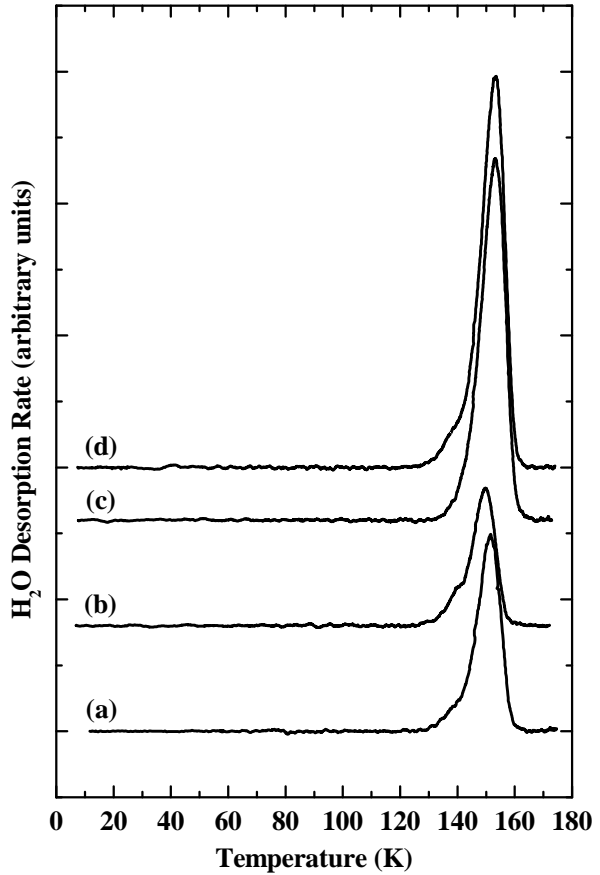}}}
{\caption{TPD spectra for H$_{{\rm 2}}$O ice films prepared (a) by 20 L
exposure at 10 K, (b) by 20 L exposure at 100 K, (c) 80 L exposure at 130 K,
and (d) 20 L exposure at 100 K followed by 20 L exposure at 10 K. Each film
was then cooled to 10 K prior to desorption. For clarity, spectra are offset
and only shown from 100 K.
}}
\end{figure}

In some cases, and particularly at higher exposure, a slight shoulder is
observed on the leading edge of the 140 K TPD peak. Under these experimental
conditions, the phase transition between amorphous and crystalline ice is
known to occur between 120 and 140 K (Sack \& Baragiola 1993). It might be
expected therefore that all the desorbing ice should be in the crystalline
phase by the time desorption occurs. However, given the relatively long
timescale for this phase change (Dohn\'{a}lek et al., 2000), it is likely
that some desorption will occur from residual amorphous phase that has yet
to undergo the phase change. The TPD peak of the amorphous ice will occur at
a lower temperature than the main desorption peak since the vapour pressure
of the amorphous ice is around three times greater than that of crystalline
ice (Kouchi 1987). This behaviour is clearly illustrated in Fig. 3. Traces
(a) and (b) show the TPD spectra from H$_{{\rm 2}}$O films of equivalent
surface exposure, deposited at two temperatures, 10 and 100 K, producing
high-density amorphous ice and low-density amorphous ice respectively. The
films were cooled to 10 K and then treated as described in \S 2. Their TPD
spectra are essentially equivalent, showing the amorphous ice desorption
shoulder on the leading edge, which gradually flattens out as the
crystallisation phase change dominates, and finally the major TPD peak,
corresponding to desorption of the crystalline phase. Two further films,
with double the surface coverage of the amorphous films, were also
investigated. Trace (c) shows results for an H$_{{\rm 2}}$O film that was
grown at 130 K then cooled to 10 K and warmed. The TPD spectrum from this
ice film is narrower, with the onset of desorption occurring slightly later
than the other spectra, and no evidence of the amorphous shoulder can be
seen. These features are attributed to the purely crystalline nature of this
ice film. Finally trace (d) in this spectrum, was produced by depositing
equal amounts of H$_{{\rm 2}}$O at 100 and then 10 K. The total surface
coverage in (d) is therefore equivalent to (c). The TPD trace for this
sample shows evidence of the amorphous phase desorption, but the phase
change appears to be much faster than in (a) or (b), and spectrum is
dominated by desorption from the crystalline phase. Similar results have
been found in previous studies of ice desorption from various substrates,
where the rate of crystallisation has also been determined (Speedy et al.
1996; Smith et al. 1996).

\bigskip

A kinetic model was used to evaluate the surface binding energy,
$E_{des}$, and the pre-exponential factor, $A$, for the crystalline H$_{{\rm
2}}$O ice system. To model the kinetics of the bulk ice desorption, a simple
reaction scheme was constructed to describe the desorption process

\begin{equation}
\label{eq5}
H_{2} O(s){\mathop { \to} \limits^{k_{d}} } H_{2} O(g)
\end{equation}

\begin{equation}
\label{eq6}
H_{2} O(g){\mathop { \to} \limits^{k_{p}} } pump
\end{equation}

\noindent
where the concentration of $H_{2}O(s)$ is equivalent to $N_{S}$, the number of
adsorbed molecules on the surface, and $k_{p}$ is the pumping coefficient of
H$_{{\rm 2}}$O in this system. A zero order rate equation, based on equation
(\ref{eq1}), was used to describe the desorption of molecules into the gas phase

\begin{equation}
\label{eq7}
{\frac{{d{\left[ {H_{2} O(g)} \right]}}}{{dt}}}\,\, = k_{d}
\end{equation}

The rate coefficient, $k_{d}$, is equivalent to that described in equation
(\ref{eq2}). A similar first order rate equation was used to describe equation (\ref{eq6})

\begin{equation}
\label{eq7} {\frac{{d{\left[ {H_{2} O(g)} \right]}}}{{dt}}}\,\, =
-k_{d}{\left[ {H_{2} O(g)} \right]}
\end{equation}

The TPD spectrum then represents the temporal evolution of these coupled
differential equations as the system temperature is raised in a linear
manner. Using an initial H$_{{\rm 2}}$O ice surface density of 2.44$\times
$10$^{{\rm 1}{\rm 6}}$ molecules cm$^{{\rm -} {\rm 2}}$ (corresponding to a
20 L exposure), an initial temperature of 10 K, and a linear temperature
ramp of 0.02 K s$^{{\rm -} {\rm 1}}$, model TPD spectra were calculated
using a simple stochastic integration package$^{{\rm \ast
}}$\footnote{$^{{\rm \ast} }$ Chemical Kinetics Simulator, Version 1.0, IBM,
IBM Almaden Research Centre, 1995. Further information may be obtained from
the CKS website at http://www.almaden.ibm.com/st/msim/ckspage.html} (Houle
\& Hinsberg 1995) and compared with the empirical data. The unknown
parameters in the model, the surface binding energy, $E_{des}$, and the
pre-exponential for desorption, $A$, were then varied to best reproduce the
empirical data. This was achieved for an $A$ value of 10$^{{\rm 3}{\rm 0}{\rm
}{\rm \pm} {\rm 2}}$ molecules cm$^{{\rm -} {\rm 2}}$ s$^{{\rm -} {\rm 1}}$
and an $E_{des}$ value of 5773 $\pm $60 K. Fig. 4 shows the comparison of the
model results with the empirical data.

\begin{figure}[!ht]
{\hbox to 0.45\textwidth{\epsfxsize=0.45\textwidth 
\epsfbox{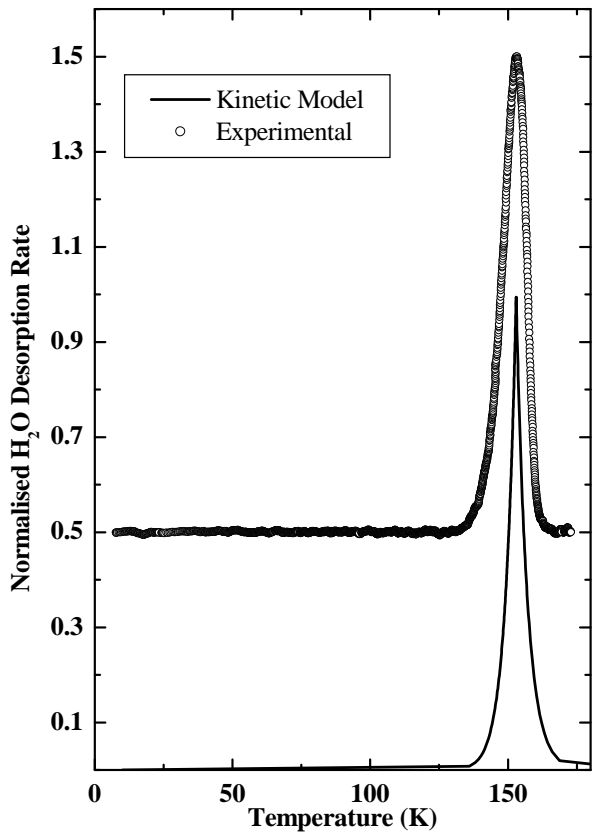}}}
{\caption{A comparison between experimental and modelled TPD spectra. See
text for discussion. }}
\end{figure}

It is much more complicated to evaluate the $A$ and $E_{des{\rm} }$values of the
amorphous phase as the desorption and crystallisation processes are
occurring simultaneously. To evaluate the gas phase H$_{{\rm 2}}$O
concentration as measured in the experiment, it is necessary to couple
equations describing the amorphous ice desorption rate and the
crystallisation rate to equations (\ref{eq6}) and (\ref{eq7}). This work is currently
ongoing, and will be the subject of a later paper. From the shapes and
behaviour of the TPD curves in Fig.1 and Fig.3 however, it is clear that
that the desorption mechanism for the amorphous ice is identical to that of
the crystalline ice. Previous studies have drawn the same conclusion
(Dohn\'{a}lek et al., 1999, 2000). Initial results from our model suggest
that the $A $factor for amorphous ice is of the same order of magnitude, and
that $E_{des}$ is a few hundred K lower, than that of crystalline ice.

\section{ASTROPHYSICAL IMPLICATIONS}

\begin{table*}[!ht]
\begin{tabular}
{|p{125pt}|p{62pt}|p{70pt}|p{68pt}|p{66pt}|} (i) Crystalline Ice & & & & \\ 
& & & & \\\hline\hline Substrate &
Au$^{{\rm a}}$ & Ru (001) / \par Au (111)$^{{\rm b}}$ &
Sapphire$^{{\rm c}}$& CsI$^{{\rm d}}$ \\
\hline E$_{{\rm d}{\rm e}{\rm s}}$ (K) \par & \textbf{5773 $\pm
$60}& 5803 $\pm $96& 5989&
5070 $\pm $50 \\
\hline A $\times $10$^{{\rm 3}{\rm 0}}$
\par (molecules cm$^{{\rm -} {\rm 2}}$ s$^{{\rm -} {\rm 1}}$) \par
& \textbf{1} & 4.58& 2.8&
*2$\times $10$^{{\rm 1}{\rm 2}}$s$^{{\rm -} {\rm 1}}$ \\
\hline
 m & \textbf{0}& 0& 0& 1 \\
\hline\hline
\end{tabular}
\end{table*}

\begin{table*}[!ht]
\begin{tabular}
{|p{125pt}|p{62pt}|p{70pt}|p{68pt}|p{66pt}|}
(ii) Amorphous Ice & & & & \\
& & & & \\
\hline\hline
Substrate& 
Au$^{{\rm a}}$& 
Ru (001) / \par Au (111)$^{{\rm b}}$& 
Sapphire$^{{\rm c}}$& 
CsI$^{{\rm d}}$ \\
\hline
E$_{{\rm d}{\rm e}{\rm s}}$ (K) \par & 
\textbf{5773 $\pm $60}& 
5803 $\pm $96& 
5989& 
5070 $\pm $50 \\
\hline
A $\times $10$^{{\rm 3}{\rm 0}}$ \par (molecules cm$^{{\rm -} {\rm 2}}$ s$^{{\rm
 -} {\rm 1}}$) \par & 
\textbf{1} & 
4.58& 
2.8& 
*2$\times $10$^{{\rm 1}{\rm 2}}$s$^{{\rm -} {\rm 1}}$ \\
\hline
m& 
\textbf{0}& 
0& 
0& 
1 \\
\hline\hline
\end{tabular}
\caption{A comparison surface binding energy, $E_{des}$, and pre-exponential
factor, $A$, measurements for (i) crystalline ice and (ii) amorphous ice. 
$^{*}$ units in s$^{-1}$ not molecules cm$^{-2}$ s$^{-1}$ as reaction was assumed to be first and not zero order. 
(a) This work, (b) Speedy et al. (1996), (c) Haynes et al. (1992), (d) Sandford \& Allamandola (1988). }
\end{table*}

The results obtained here compare favourably with previous TPD measurements
of H$_{{\rm 2}}$O desorption under UHV conditions, from a range of surfaces,
(Kay et al$.$ 1989; Haynes Tro \& George 1992; Smith Huang \& Kay 1997;
Chakarov \& Kasemo 1998; Dohn\'{a}lek et al$.$ 1999, 2000). When values for
$E_{des}$ and $A$ are provided, the data are summarised in Table 1, where a
comparison is also made with the spectroscopic results of Sandford \&
Allamandola (1988). With the exception of the latter results, the table
clearly shows that for both ordered and disordered hydrophobic surfaces, the
desorption energy and rate constant are independent of the substrate. In
each case, the desorption kinetics resemble zero order rather than first
order reaction kinetics.

\begin{table*}[!ht]
\begin{tabular}{cllllllll}
\hline\hline
\raisebox{-3.00ex}[0cm][0cm]{\textbf{Grain T} \par \textbf{(K)}}& 
\multicolumn{8}{c}{Half-life, $t_{1 / 2}$, (yr.)}  \\
\cline{2-9} 
 & 
\multicolumn{4}{c}{Zero order desorption kinetics, i.e. $t_{{\raise0.5ex\hbox{$\scriptstyle {1}$}\kern-0.1em/\kern-0.15em\lower0.25ex\hbox{$\scriptstyle {2}$}}} = {\frac{{N_{S,0}} }{{2k_{d}} }}$} & 
\multicolumn{4}{c}{First order desorption kinetics, i.e. $t_{{\raise0.5ex\hbox{$\scriptstyle {1}$}\kern-0.1em/\kern-0.15em\lower0.25ex\hbox{$\scriptstyle {2}$}}} = {\frac{{\ln 2}}{{k_{d}} }}$ }   \\
& \multicolumn{4}{c}{where $N_{s,0} $=1.15$\times $10$^{{\rm 1}{\rm 7}}$ molecules cm$^{{\rm -} {\rm 2}}$} & \multicolumn{4}{c}{} \\
\cline{2-9} 
 & 
\multicolumn{2}{c}{H$_{{\rm 2}}$O on crystalline ice} & 
\multicolumn{2}{c}{H$_{{\rm 2}}$O on amorphous ice} & 
\multicolumn{2}{c}{H$_{{\rm 2}}$O on crystalline ice} & 
\multicolumn{2}{c|}{H$_{{\rm 2}}$O on amorphous ice}  \\
 & \multicolumn{1}{c}{(a)} & \multicolumn{1}{c}{(b)} & \multicolumn{1}{c}{(a)} & \multicolumn{1}{c}{(b)} & \multicolumn{1}{c}{(c)} & \multicolumn{1}{c}{(d)} & \multicolumn{1}{c}{(c)} & \multicolumn{1}{c}{(d)} \\
\hline
\textbf{10}& 
\textbf{8.8$\times $10}$^{{\rm 2}{\rm 3}{\rm 1}}$\textbf{}& 
\textit{1.3$\times $10}$^{200}$\textit{}& 
\textbf{6$\times $10}$^{{\rm 2}{\rm 2}{\rm 4}}$\textbf{}& 
\textit{8.9$\times $10}$^{188}$\textit{}& 
4.6$\times $10$^{{\rm 2}{\rm 3}{\rm 2}}$& 
\textit{6.8$\times $10}$^{200}$\textit{}& 
3.8$\times $10$^{{\rm 2}{\rm 2}{\rm 5}}$& 
\textit{5.7$\times $10}$^{189}$\textit{} \\
\hline
\textbf{20}& 
\textbf{8.6$\times $10}$^{{\rm 1}{\rm 0}{\rm 5}}$\textbf{}& 
\textit{1$\times $10}$^{90}$\textit{}& 
\textbf{2$\times $10}$^{{\rm 1}{\rm 0}{\rm 2}}$\textbf{}& 
\textit{2.5$\times $10}$^{84}$\textit{}& 
4.5$\times $10$^{{\rm 1}{\rm 0}{\rm 6}}$& 
\textit{5.5$\times $10}$^{90}$\textit{}& 
1.3$\times $10$^{{\rm 1}{\rm 0}{\rm 3}}$& 
\textit{1.6$\times $10}$^{85}$\textit{} \\
\hline
\textbf{30}& 
\textbf{8.5$\times $10}$^{{\rm 6}{\rm 3}}$\textbf{}& 
\textit{2.1$\times $10}$^{53}$\textit{}& 
\textbf{3$\times $10}$^{{\rm 6}{\rm 1}}$\textbf{}& 
\textit{3.5$\times $10}$^{49}$\textit{}& 
4.5$\times $10$^{{\rm 6}{\rm 4}}$& 
\textit{1.1$\times $10}$^{54}$\textit{}& 
2$\times $10$^{{\rm 6}{\rm 2}}$& 
\textit{2.2$\times $10}$^{50}$\textit{} \\
\hline
\textbf{40}& 
\textbf{8.4$\times $10}$^{{\rm 4}{\rm 2}}$\textbf{}& 
\textit{9.3$\times $10}$^{34}$\textit{}& 
\textbf{1.2$\times $10}$^{{\rm 4}{\rm 1}}$\textbf{}& 
\textit{1.3$\times $10}$^{32}$\textit{}& 
4.4$\times $10$^{{\rm 4}{\rm 3}}$& 
\textit{4.9$\times $10}$^{35}$\textit{}& 
7.6$\times $10$^{{\rm 4}{\rm 1}}$& 
\textit{8.3$\times $10}$^{32}$\textit{} \\
\hline
\textbf{50}& 
\textbf{2.1$\times $10}$^{{\rm 3}{\rm 0}}$\textbf{}& 
\textit{9.1$\times $10}$^{23}$\textit{}& 
\textbf{6.7$\times $10}$^{{\rm 2}{\rm 8}}$\textbf{}& 
\textit{4.5$\times $10}$^{21}$\textit{}& 
1.1$\times $10$^{{\rm 3}{\rm 1}}$& 
\textit{4.8$\times $10}$^{24}$\textit{}& 
4.3$\times $10$^{{\rm 2}{\rm 9}}$& 
\textit{2.9$\times $10}$^{22}$\textit{} \\
\hline
\textbf{60}& 
\textbf{8.4$\times $10}$^{{\rm 2}{\rm 1}}$\textbf{}& 
\textit{4.2$\times $10}$^{16}$\textit{}& 
\textbf{4.6$\times $10}$^{{\rm 2}{\rm 0}}$\textbf{}& 
\textit{4.9$\times $10}$^{14}$\textit{}& 
4.4$\times $10$^{{\rm 2}{\rm 2}}$& 
\textit{2.2$\times $10}$^{17}$\textit{}& 
2.9$\times $10$^{{\rm 2}{\rm 1}}$& 
\textit{3.1$\times $10}$^{15}$\textit{} \\
\hline
\textbf{70}& 
\textbf{8.4$\times $10}$^{{\rm 1}{\rm 5}}$\textbf{}& 
\textit{2.4$\times $10}$^{11}$\textit{}& 
\textbf{6.7$\times $10}$^{{\rm 1}{\rm 4}}$\textbf{}& 
\textit{5.1$\times $10}$^{9}$\textit{}& 
4.4$\times $10$^{{\rm 1}{\rm 6}}$& 
\textit{1.3$\times $10}$^{12}$\textit{}& 
4.3$\times $10$^{{\rm 1}{\rm 5}}$& 
\textit{3.3$\times $10}$^{10}$\textit{} \\
\hline
\textbf{80}& 
\textbf{2.7$\times $10}$^{{\rm 1}{\rm 1}}$\textbf{}& 
\textit{2.8$\times $10}$^{7}$\textit{}& 
\textbf{2.8$\times $10}$^{{\rm 1}{\rm 0}}$\textbf{}& 
\textit{9.4$\times $10}$^{5}$\textit{}& 
1.4$\times $10$^{{\rm 1}{\rm 2}}$& 
\textit{1.5$\times $10}$^{8}$\textit{}& 
1.8$\times $10$^{{\rm 1}{\rm 1}}$& 
\textit{6.1$\times $10}$^{6}$\textit{} \\
\hline
\textbf{90}& 
\textbf{8.4$\times $10}$^{{\rm 7}}$\textbf{}& 
\textit{2.4$\times $10}$^{4}$\textit{}& 
\textbf{1.1$\times $10}$^{{\rm 7}}$\textbf{}& 
\textit{1.2$\times $10}$^{3}$\textit{}& 
4.4$\times $10$^{{\rm 8}}$& 
\textit{1.3$\times $10}$^{5}$\textit{}& 
7.2$\times $10$^{{\rm 7}}$& 
\textit{7.5$\times $10}$^{3}$\textit{} \\
\hline
\textbf{100}& 
\textbf{1.3$\times $10}$^{{\rm 5}}$\textbf{}& 
\textit{8.7$\times $10}$^{1}$\textit{}& 
\textbf{2.1$\times $10}$^{{\rm 4}}$\textbf{}& 
\textit{5.6$\times $10}$^{0}$\textit{}& 
7$\times $10$^{{\rm 5}}$& 
\textit{4.6$\times $10}$^{2}$\textit{}& 
1.4$\times $10$^{{\rm 5}}$& 
\textit{3.6$\times $10}$^{1}$\textit{} \\
\hline
\textbf{110}& 
\textbf{6.8$\times $10}$^{{\rm 2}}$\textbf{}& 
\textit{8.7$\times $10}$^{ - 1}$\textit{}& 
\textbf{1.3$\times $10}$^{{\rm 2}}$\textbf{}& 
\textit{7$\times $10}$^{ - 2}$\textit{}& 
3.6$\times $10$^{{\rm 3}}$& 
\textit{4.6$\times $10}$^{0}$\textit{}& 
8.1$\times $10$^{{\rm 2}}$& 
\textit{4.5$\times $10}$^{ - 1}$\textit{} \\
\hline
\textbf{120}& 
\textbf{8.4$\times $10}$^{{\rm 0}}$\textbf{}& 
\textit{1.8$\times $10}$^{ - 3}$\textit{}& 
\textbf{1.8$\times $10}$^{{\rm 0}}$\textbf{}& 
\textit{1.9$\times $10}$^{ - 2}$\textit{}& 
4.1$\times $10$^{{\rm 1}}$& 
\textit{9.8$\times $10}$^{ - 2}$\textit{}& 
1.1$\times $10$^{{\rm 1}}$& 
\textit{1.2$\times $10}$^{ - 2}$\textit{} \\
\hline
\textbf{130}& 
\textbf{2$\times $10}$^{{\rm -} {\rm 1}}$\textbf{}& 
\textit{7.2$\times $10}$^{ - 4}$\textit{}& 
\textbf{4.7$\times $10}$^{{\rm -} {\rm 2}}$\textbf{}& 
\textit{8.3$\times $10}$^{ - 5}$\textit{}& 
1.1$\times $10$^{{\rm 0}}$& 
\textit{3.8$\times $10}$^{ - 3}$\textit{}& 
3.1$\times $10$^{{\rm -} {\rm 1}}$& 
\textit{5.3$\times $10}$^{ - 4}$\textit{} \\
\hline
\textbf{140}& 
\textbf{8.4$\times $10}$^{{\rm -} {\rm 3}}$\textbf{}& 
\textit{5.9$\times $10}$^{ - 6}$\textit{}& 
\textbf{2.1$\times $10}$^{{\rm -} {\rm 3}}$\textbf{}& 
\textit{4.5$\times $10}$^{ - 5}$\textit{}& 
4.2$\times $10$^{{\rm -} {\rm 2}}$& 
\textit{2.3$\times $10}$^{ - 4}$\textit{}& 
1.4$\times $10-2& 
\textit{3.8$\times $10}$^{ - 5}$\textit{} \\
\hline
\textbf{150}& 
\textbf{5.3$\times $10}$^{{\rm -} {\rm 4}}$\textbf{}& 
\textit{4$\times $10}$^{ - 6}$\textit{}& 
\textbf{1.5$\times $10}$^{{\rm -} {\rm 4}}$\textbf{}& 
\textit{6$\times $10}$^{ - 7}$\textit{}& 
2.8$\times $10$^{{\rm -} {\rm 3}}$& 
\textit{2.1$\times $10}$^{ - 5}$\textit{}& 
9.4$\times $10$^{{\rm -} {\rm 4}}$& 
\textit{3.8$\times $10}$^{ - 6}$\textit{} \\
\hline\hline
\end{tabular}
\caption{The half-life, $t_{1 / 2}$, of H$_{{\rm 2}}$O molecules on H$_{{\rm
2}}$O ice surfaces as a function of temperature under zero and first order
kinetics. In each case, the half-life is defined in the table, and
represents the time it takes for the surface population of H$_{{\rm 2}}$O
molecules on an interstellar grain to decrease to one half of its initial
value. 
(a) $E_{des}$ and $A$ taken from this work,  (b) $E_{des}$ taken from Sandford \& Allamandola (1988), $A$ taken from this work, (c) $E_{des}$ taken from this work, $A$ taken from Sandford \& Allamandola (1988), (d) $E_{des}$ and $A$ taken from Sandford \& Allamandola (1988)}
\end{table*}

None of the studies mentioned in Table 1 have identified any differences
between the desorption of H$_{{\rm 2}}$O from H$_{{\rm 2}}$O layers and the
desorption of H$_{{\rm 2}}$O from the underlying substrate. This indicates
that H$_{{\rm 2}}$O is only weakly bound to hydrophobic substrates, and that
the adsorbate-substrate binding energy is comparable to the
adsorbate-adsorbate hydrogen bonding. In astrophysical terms, this indicates
that the same desorption mechanism and desorption energy can be used to
model H$_{{\rm 2}}$O desorption from any hydrophobic surface, (e.g. silicon,
graphite). Similar desorption studies on hydrophilic surfaces (such as metal
oxides, quartz, and silicates) indicate that H$_{{\rm 2}}$O bonds chemically
to such surfaces. In such cases, the H$_{{\rm 2}}$O-substrate system
exhibits alternative desorption kinetics, with even higher activation
barriers (for example\textit{} Stirniman et al. 1996; Trakhtenberg et al$.$ 1997; Hudson
et al. 2001). However, these systems do exhibit identical H$_{{\rm
2}}$O-H$_{{\rm 2}}$O desorption processes in the regime where the ice
surface is substrate independent ( i.e. the ice layer is thick enough that
it is no longer influenced by the underlying substrate). Consequently, in
ISM regions where relatively thick ice layers are formed, or the grain
surface is hydrophobic, the underlying structure of the cosmic grain is not
significant. Elsewhere, the grain material may influence the H$_{{\rm 2}}$O
desorption mechanism. It will not be possible to say conclusively which
H$_{{\rm 2}}$O desorption kinetics will dominate until the nature of
interstellar grains has been clearly established.

\bigskip

As Table 1 shows, the results obtained in this study differ significantly
from those obtained by Sandford \& Allamandola 1988. Firstly, the
$E_{des}$ differs by \textit{ca.} 800 K for both the crystalline and amorphous ice
phases, and secondly the desorption kinetics in this study are clearly zero
order and not first order. These two parameters are both significant when
modelling interstellar environments, particularly when calculating the
residence time of H$_{{\rm 2}}$O on interstellar grains, or establishing the
ratio of solid to gas phase material.

\bigskip

Why do the two methods give such different results? The discrepancies must
be related to the different assumptions that are made in interpreting the
data under each set of experimental conditions. The most significant
difference between the two experiments is that the spectroscopic data are
taken under isothermal conditions at base pressures of \textit{ca.} 2$\times $10$^{{\rm
-} {\rm 8}}$ Torr, whereas the TPD spectra are recorded in dynamic
conditions at base pressures of 6$\times $10$^{{\rm -} {\rm 1}{\rm 1}}$
Torr. In the spectroscopic measurements, it was assumed that a first order
rate law applied, and the value of $E_{des{\rm} }$was determined indirectly,
by monitoring the relative change in integrated intensity of a surface IR
spectroscopic feature during isothermal desorption. Contrary to the
assumptions made in Sandford \& Allamandola's method, close to the
desorption temperature, the sticking probability of H$_{{\rm 2}}$O molecules
on the surface is no longer unity, and over the time period of the
experiment it is possible for re-adsorption to occur, both from vacuum
contaminants and desorbing gas. The evaluation of isothermal measurements is
also complicated by non-equilibrium effects, including both crystallisation
and re-adsorption. In the TPD experiments it is possible to obtain
information on the order of the reaction and $E_{des}$ directly, from the
shape and position of the TPD curve. In UHV systems designed for TPD
experiments, the pumping speed is so high that surface re-adsorption is not
relevant. Unlike the spectroscopic data, the TPD spectra clearly show that
crystallisation and desorption are occurring concurrently in the system. It
is possible however to identify the amorphous, crystalline and `phase
change' regions of the TPD spectra from each other. With careful film
preparation it is even possible to force the TPD spectrum to exhibit only
crystalline behaviour, as was seen in Fig. 3. Furthermore, Kay and
co-workers have made a number studies using TPD methods to determine the
rates of crystallisation and desorption when they are occurring at the same
time, and have found no significant difference between the desorption rates
they measure isothermally or dynamically (Speedy et al. 1996; Smith et al.
1996; Dohn\'{a}lek et al. 1999, 2000). Finally the experimental conditions
under which the TPD spectra were obtained most closely resemble those in the
interstellar medium. We therefore believe that these data represent the
parameters for H$_{{\rm 2}}$O that should be employed in astronomical
models.

\bigskip

Table 2 shows the effect these new parameters have on the 'residence times'
of H$_{{\rm 2}}$O molecules on H$_{{\rm 2}}$O ices and hydrophobic surfaces.
It is not possible to define a residence time for a zero order process that
is directly comparable (with the same units) with a first order process.
Consequently, the results are presented here in terms of the half-life of
the H$_{{\rm 2}}$O surface population, rather than the 'residence time,
1/$k_{d}$' that is commonly used in the astronomical literature. The
respective half-life values are defined in the table. It is important to
note that the half-life of a zero order process also depends on the surface
concentration of H$_{{\rm 2}}$O molecules. In these calculations this value
was fixed at 1.15$\times $10$^{{\rm 1}{\rm 7}}$ molecules cm$^{{\rm -} {\rm
2}}$, equivalent to around 100 monolayers of H$_{{\rm 2}}$O ice, i.e. the
number of H$_{{\rm 2}}$O layers expected to accrete on a interstellar grain
during the lifetime of a dense cloud (Hasegawa \& Herbst 1993). The
half-lives calculated using the zero order $A $and $E_{des}$ values obtained in
this work are given in column one (crystalline ice) and three (amorphous
ice).The results are also compared to half-life calculations using the data
from Sandford \& Allamandola (1993). The half-lives calculated using
Sandford \& Allamandola's first order $A $and $E_{des}$ values are given in
columns six (crystalline ice) and eight (amorphous ice). From Table 2 it is
immediately clear that H$_{{\rm 2}}$O can remain on the grain surface at
significantly higher temperatures than has previously been assumed,
typically around 110 - 120 K rather than 90 - 100 K.

\section{CONCLUSION}

The surface binding energy and desorption kinetics of amorphous and
crystalline H$_{{\rm 2}}$O ice have been studied under conditions similar to
those found in denser regions of interstellar clouds. These data have been
used to calculate the residence times of H$_{{\rm 2}}$O on H$_{{\rm 2}}$O
and hydrophobic surfaces as a function of temperature. The results imply
that, in the dense interstellar medium, thermal desorption of H$_{{\rm 2}}$O
ice will occur at significantly higher temperatures, and different rates,
than previously had been assumed. These data are of fundamental importance
in the chemical modelling of many astrophysical environments. The effects
may be particularly pronounced in so-called hot cores, which are very dense
clumps of gas, remnants of a collapsing cloud that formed a massive star.
Irradiation of the clumps heats the cores to temperatures in the range 100 -
300 K, when the ices evaporate, populating the gas with water and other
trace molecules (cf. Millar 1993). The temperature of the evaporation may be
crucial to relating the observed chemistry of hot cores to the implied rate
of warming of the central star (Viti and Williams 1999).The results provide
the astronomical community with reliable thermodynamic data pertaining to
the desorption of H$_{{\rm 2}}$O under interstellar conditions.

\bigskip

\section*{ACKNOWLEDGEMENTS}

The authors thank PPARC for their financial support, without which these
experiments would not have been possible.

\end{document}